\documentclass[twocolumn]{aastex631}
\DeclareUnicodeCharacter{200B}{}
\usepackage{color}
\usepackage[utf8]{inputenc}
\usepackage[T1]{fontenc}
\usepackage{amsmath}

\shortauthors{Zhu, Kewley \& Sutherland}

\begin{document}
	
\title{A Theoretical Three-Dimensional Diagram to Separate Star Formation, Active Galactic Nuclei, and Shocks in Galaxies}

\author[0000-0002-1333-147X]{Peixin Zhu}
\affiliation{Research School of Astronomy and Astrophysics, Australian National University, Australia}
\affiliation{ARC Centre of Excellence for All Sky Astrophysics in 3 Dimensions (ASTRO 3D), Australia}
\affiliation{Center for Astrophysics $|$ Harvard \& Smithsonian, 60 Garden Street, Cambridge, MA 02138, USA}

\author[0000-0001-8152-3943]{Lisa J. Kewley}
\affiliation{Research School of Astronomy and Astrophysics, Australian National University, Australia}
\affiliation{ARC Centre of Excellence for All Sky Astrophysics in 3 Dimensions (ASTRO 3D), Australia}
\affiliation{Center for Astrophysics $|$ Harvard \& Smithsonian, 60 Garden Street, Cambridge, MA 02138, USA}

\author[0000-0002-6620-7421]{Ralph S. Sutherland}
\affiliation{Research School of Astronomy and Astrophysics, Australian National University, Australia}

\author[0000-0002-3247-5321]{Kathryn Grasha}
\affiliation{Research School of Astronomy and Astrophysics, Australian National University, Australia}
\affiliation{ARC Centre of Excellence for All Sky Astrophysics in 3 Dimensions (ASTRO 3D), Australia}
\affiliation{ARC DECRA Fellow}

\email{peixin.zhu@cfa.harvard.edu}

\begin{abstract}

The excitation sources in galaxies are frequently mixed due to AGN and stellar feedback, including star formation, active galactic nuclei (AGNs), and shock excitation. Disentangling the star formation, AGN, and shocks in galaxy integral-field spectra (IFU) at optical wavelengths is crucial to expanding the galaxy sample for AGN and stellar feedback studies, given the lack of multiwavelength observations for most of the galaxies that are observed in optical wavelengths. Previous methods to address this issue either have a limited application range or are highly uncertain in separating AGN from shock excitation \citep{dagostino_new_2019,johnston_beyond_2023}. Here, we propose a theoretical three-dimensional (3D) diagram. This theoretical 3D diagram overcomes the limitations of previous methods and can simultaneously separate star formation, AGNs, and shocks in active galaxies. Along with the separation, the new theoretical 3D diagram also constrains the gas metallicity, ionization parameter, and gas pressure within the galaxy. By applying the Very Large Telescope (VLT)/MUSE IFU data and the Wide Field Spectrograph IFU data for NGC\,5728 on the theoretical 3D diagram, we find a star-forming ring surrounding the galaxy center with a projected radius of $\sim1\,$kpc in the sky plane, an AGN ionized-bicone extended up to $\sim2\,$kpc from the nuclear center, and a fast shock dominated disk region at the base of the AGN outflow, which is likely associated with a nuclear accretion disk or a result of jet-ISM interaction. The theoretical 3D diagram opens a new window to study the interplay among star formation, AGN, and shocks in active galaxies.


\end{abstract}


\section{Introduction}

Multiple excitation sources are known to exist in active galaxies, including bright O and B stars and Wolf-Rayet Stars \citep[e.g.][]{lutz_what_1996,kewley_theoretical_2001}, actively accreting black holes \citep[e.g.][]{osterbrock_astrophysics_1989,groves_dusty_2004-1}, shocks \citep[e.g.][]{dopita_spectral_1995,rich_galaxy-wide_2011,juneau_black_2022}, and diffuse ionized gas \citep{reynolds_measurement_1984,haffner_warm_2009,belfiore_sdss_2016,poetrodjojo_effects_2019,mannucci_diffuse_2021}. All of these mechanisms contribute to the emission lines in the galaxy spectrum. A major challenge in galaxy studies that rely on emission lines is properly disengaging the mixture of different excitation sources.

Studies of star-forming galaxies show that AGN can contribute up to 40\% of the H$\alpha$ emission in the galaxies \citep{davies_starburstagn_2014,davies_starburstagn_2014-1,cairos_warm_2022}. Therefore the mixing of star formation and AGN can mislead the emission-line-based estimation for star formation rates (SFRs) and the bolometric luminosity of AGN, resulting in false conclusions on the relation between AGN activity and the host galaxy properties. 

{\color{black}Shocks introduce an additional complexity to the excitation mixture in galaxies. They can be generated by feedback from AGNs \citep{blustin_nature_2005, king_powerful_2015, laha_ionized_2021, juneau_black_2022} or from massive stars \citep{schwartz_t_1975, reipurth_giant_1997, dopita_effects_2017}, particularly at the interface between energetic outflows and the surrounding interstellar medium. Identifying and separating the contribution of shocks is therefore essential for spatially resolved studies to understand AGN feedback and the star formation quenching mechanism.

However, despite their physical significance, shocks are frequently neglected in excitation mechanism separation studies. This is largely because shocks are challenging to distinguish from AGN ionization in two-dimensional (2D) optical diagnostic diagrams \citep[e.g.,][]{baldwin_classification_1981, veilleux_spectral_1987-1, kewley_host_2006}.} Both observational studies \citep{rich_galaxy_2015, mortazavi_dynamics_2019} and theoretical modeling \citep{allen_mappings_2008, sutherland_effects_2017, kewley_understanding_2019} have shown that shocks can mimic AGN emission lines on the 2D standard optical diagnostic diagrams, making it difficult to distinguish shocks from the AGN using 2D diagrams alone (see Figure 11 in \citet{kewley_understanding_2019}). 

Separating the mixture of excitation sources within a galaxy is made possible by wide-area and high-spatial-resolution integral field unit spectroscopy (IFU). Unlike the single-aperture spectroscopy from which only the galaxy center spectrum is obtained, IFU provides multiple spectra for a single galaxy, spanning from the galaxy center region to the outer regions. Combined with the optical diagnostic diagrams, IFU data can reveal the spatial distribution of the dominant excitation source in a galaxy \citep[e.g.][]{davies_starburstagn_2014-1,davies_starburstagn_2014,ho_sami_2014,dagostino_starburstagn_2018}. Large integral field galaxy surveys further enable statistical studies on spatially resolved galaxy properties. The completed large optical integral field surveys including the Calar Alto Legacy Integral Field Area (CALIFA, \citet{sanchez_califa_2012}) survey, the Sydney Australian Astronomical Observatory Multi-object Integral Field Spectrograph Galaxy Survey (SAMI, \citet{allen_sami_2015}), and the Sloan Digital Sky Survey IV Mapping Nearby Galaxies at Apache Point Observatory survey (MaNGA, \citet{bundy_overview_2015}). 

With IFU data, many attempts have been made to separate the mixed excitation sources. In the case where only two excitation sources (star formation and AGN, or star formation and shocks) are present, separation can be performed on 2D optical diagnostic diagrams \citep[e.g.][]{rich_galaxy-wide_2011,ho_sami_2014,davies_starburstagn_2014,medling_shocked_2015,durre_agn_2018,dagostino_starburstagn_2018}. 

For example, in the case where only star formation and AGN are present, \citet{davies_starburstagn_2014-1,davies_starburstagn_2014,dagostino_starburstagn_2018} perform star formation and AGN separation for NGC\,7130, NGC\,5728, NGC\,7582, and NGC\,1068 using the Baldwin-Phillips-Terlevich (BPT) diagrams \citep{baldwin_classification_1981,veilleux_spectral_1987-1}. They define a star formation-AGN mixing sequence for each galaxy on the BPT diagram and assign two basis points as 100\% AGN point and 100\% star formation point at each end of the sequence. For each spaxel along the star formation-AGN mixing sequence, a fractional contribution from star formation and AGN is assigned based on the relative distances to the two basis points. Further examples of separating star formation and shocks on 2D diagrams can be found in \citet{rich_galaxy-wide_2011,ho_sami_2014,medling_shocked_2015,alatalo_shocked_2016}.

However, separating star formation, AGN, and shocks simultaneously is extremely challenging and usually requires including gas kinematic information of galaxies. 

The first attempt to perform star formation, AGN, and shock separation on a three-dimensional (3D) diagram is performed by \citet{dagostino_comparison_2019,dagostino_new_2019}. They built an empirical 3D diagram using an emission line ratio function (ELR), the velocity dispersion of each Gaussian component, and the radial distance to the galaxy center as three axes. The ELR function, calculated from Equation~\ref{equ:1}, characterizes the location of a spaxel along the star formation-AGN mixing sequence on the BPT diagram, with the 100\% AGN point having ELR=1 and the 100\% star formation point having ELR=0. On this 3D diagram, the IFU data of NGC1068 from the Siding Spring Southern Seyfert Spectroscopic Snapshot Survey (S7, \citet{dopita_probing_2015-1,thomas_probing_2017}) is distributed into two separated mixing sequences: a star formation-AGN mixing sequence and a star formation-shocks mixing sequence. By selecting three basis points as the 100\% basis points for star formation, AGN, and shocks and calculating the distance between each spaxel to each basis point, they obtained the fractional contribution from every mechanism for each spaxel based on its relative distances to the three basis points. 

\begin{small}
\begin{equation}
\begin{aligned}
    \mathrm{ELR\,function} &= \frac{\log(\mathrm{[N~II]/H}\alpha) - \min_{\log(\mathrm{[N~II]/H}\alpha)}}{\max_{\log(\mathrm{[N~II]/H}\alpha)} - \min_{\log(\mathrm{[N~II]/H}\alpha)}} \\
    &\quad \times \frac{\log(\mathrm{[O~III]/H}\beta) - \min_{\log(\mathrm{[O~III]/H}\beta)}}{\max_{\log(\mathrm{[O~III]/H}\beta)} - \min_{\log(\mathrm{[O~III]/H}\beta)}}
\end{aligned}
\end{equation}\label{equ:1}
\end{small}


Despite the success of the application of the empirical 3D diagram on NGC\,1068, the empirical 3D diagram relies heavily on an evenly distributed ELR function to obtain two well-separate sequences on the 3D diagram, which are only satisfied by galaxies whose IFU data have a complete star formation-AGN mixing sequence on the BPT diagram. After testing the empirical 3D diagram on the whole S7 sample (131 Seyfert galaxies in total), we find only 22 galaxies (17\%) exhibit two clear mixing sequences on the empirical 3D diagram (Zhu et al. 2025b, in preparation), suggesting the need for a more widely applicable method to separate star formation, AGN, and shocks simultaneously.

Following \citet{dagostino_new_2019}, \citet{johnston_beyond_2023} proposes a multidimensional diagnostic diagram to separate star formation, AGN, and shocks for galaxies in the SAMI DR3. They combine three 2D optical diagnostic diagrams with velocity dispersion and galactic radius to identify the shock ionization in galaxies. {\color{black} Among the 1996 galaxies in their sample, they find 547 galaxies that contain either AGN or shock ionization. However, 357 galaxies in their sample contain non-star-forming-like velocity dispersion which cannot distinguish between low-luminosity AGNs, shocks, and beam-smearing using their multidimensional diagram.}

An opportunity to overcome these limitations in previous methods is delivered by the latest assembly of state-of-the-art and self-consistent theoretical models for HII regions \citep{kewley_understanding_2019}, AGN narrow-line regions \citep{zhu_new_2023}, and the time-dependent shocks and precursor models \citep{sutherland_effects_2017,dopita_effects_2017}. These models can each provide reliable prediction of emission lines in the star-forming, AGN, and shock regions. These models are all calculated with the latest version of MAPPINGS V5.2 \citep{sutherland_mappings_2018} and use the same atomic data from CHIANTI v10 \citep{del_zanna_chiantiatomic_2021}, and abundances sets \citep{2017MNRAS.466.4403N}, which guarantees that the difference in the emission line prediction is only due to the different excitation sources. 

This paper uses these new theoretical models to improve upon the previous work in \citet{dagostino_comparison_2019} and \citet{johnston_beyond_2023}. We propose a theoretical 3D diagram that can simultaneously separate star formation, AGN, and shocks in a galaxy, as well as constrain the gas metallicity, ionization parameter, gas pressure, and shock velocity in corresponding regions, for the first time.

This paper is structured as follows: Section 2 describes the observational data, with theoretical models introduced in Section 3. Section 4 describes the new 3D diagram, and Section 5 presents the method to separate different mechanisms. Section 6 presents the excitation mechanisms of NGC 5728 as a test case. 
Section 7 discusses the new insights gained into galaxy NGC\,5728 as a result of the new 3D diagnostic. {\color{black}We adopt a cosmology of $H_0=70\,\rm km\,s^{-1}\,Mpc^{-1}$, $\Omega_{\Lambda}=0.7$, and $\Omega_{m}=0.3$.}

\section{Observational Data}\label{sec:data}

We utilize the optical IFU observations of NGC\,5728 obtained with both the Very Large Telescope (VLT)/MUSE and the Wide Field Spectrograph (WiFeS, \citep{dopita_wide_2007,dopita_wide_2010} on the ANU 2.3\,m telescope. 

{\color{black} NGC\,5728 serves as an ideal case for this study for three reasons: (1) NGC 5728 has been observed by two different optical IFU instruments, MUSE and WiFeS, allowing a direct comparison of how IFU resolution and data quality influence excitation source separation; (2) Previous studies have applied BPT diagrams to separate star formation and AGN in NGC\,5728 using MUSE data \citep{shin_positive_2019-1,durre_agn_2018,shimizu_multiphase_2019} and WiFeS data \citep{davies_dissecting_2016}. This enables a direct comparison with the excitation sources separation from our theoretical 3D diagram; (3) NGC 5728 has been observed across multiple wavelengths, including infrared \citep{davies_gatos_2024,zhang_galaxy_2024}, X-ray \citep{falcao_deep_2023}, and radio \citep{shin_positive_2019-1}, providing opportunities for multi-wavelength validation of our excitation sources separation.}

NGC\,5728 is observed with VLT/MUSE. MUSE has a spatial resolution of 0.2 arcsec pixel$^{-1}$. VLT/MUSE consists of 24 IFUs that cover the wavelength range 4650-9300\,\AA \citep{bacon_muse_2010} and has a FoV of $1'\times1'$. On 2016 April 3 and June 3, NGC\,5728 was observed as part of the Time Inference with MUSE in Extragalactic Rings (TIMER) project (programme ID 097.B-0640(A), PI: D. Gadotti) \citep{gadotti_time_2019}. This observation consists of 12 exposures, leading to a total exposure time of 79\,min. We retrieved the data cube from the ESO archive. The seeing size for the combined data cube is 0.66 arcsec. 

We perform single Gaussian fits to the emission lines in the MUSE IFU data using the penalized pixel-fitting routine (pPXF) module \citep{cappellari_parametric_2004,cappellari_improving_2017} incorporated in the nGIST pipeline \citep{bittner_gist_2019}\footnote{http://ascl.net/1907.025}. We use a combination of single-stellar population (SSP) modules from the MILES library \citep{vazdekis_evolutionary_2010} to remove the continuum before fitting the emission lines. In the emission-line fitting, we divide the emission lines into three groups: (i) low ionization lines [N~II],[N~I],[O~I], and [S~II]; (ii) high ionization lines [O~III], He~I, and [S~III]; (iii) Hydrogen Balmer lines H$\alpha$ and H$\beta$. In each group, the kinematic properties of velocity ($V$) and velocity dispersion ($\sigma$) are tied together in the fitting.

NGC\,5728 was also observed in the S7 survey \citep{dopita_probing_2015-1,thomas_probing_2017} on 2014 April 8 with an exposure time of 1800\,s. This observation was performed with the WiFeS on the ANU 2.3\,m telescope. The WiFeS has a field of view (FoV) of $38\times25$ arcsec$^2$ and a spatial resolution of 1 arcsec pixel$^{-1}$. WiFeS is a double-beam spectrograph, with a high spectral resolution $R=7000$ in the red (FWHM$\sim$40\,km s$^{-1}$ over 540-700\,nm) and a lower spectral resolution $R=3000$ in the blue (FWHM$\sim$100\,km s$^{-1}$ over 350-570\,nm). The seeing of this observation is 1.2\,arcsec.

The S7 survey uses LZIFU\footnote{https://github.com/hoiting/LZIFU} \citep[``Lazy-IFU''][]{ho_lzifu_2016} to perform emission-line fitting on IFU data cubes. After subtracting the continuum using the pPXF routine, LZIFU performs multi-component (1, 2, or 3) Gaussian fits to the emission lines using the Levenberg-Marquardt least-squares algorithm implemented in MPFIT \citep{markwardt_non-linear_2009}. The optimal number of Gaussians needed for each spaxel is later determined using an Artificial Neural Network (ANN, \citep{hampton_using_2017}) trained by astronomers. More details on the emission line fittings process of the S7 survey can be found in \citet{thomas_probing_2017}. We downloaded the emission line fitted data cubes from the website of the S7 survey\footnote{https://www.mso.anu.edu.au/S7/}. 

\section{theoretical models}\label{sec:model}

We use MAPPINGS version 5.2 \citep{binette_radiative_1985,sutherland_cooling_1993,dopita_new_2013,sutherland_mappings_2018} to model the HII region, AGN narrow-line region, and radiative shock in this work\footnote{{\color{black}The theoretical models used in this study will be shared upon request to the author.}}, noting the differences between MAPPINGS and CLOUDY \citep[e.g.][]{ferland_2017_2017} in optical emission line ratios prediction for [N~II]$\,\lambda6584$/H$\alpha$ and [O~III]$\,\lambda5007$/H$\beta$ are within $\sim0.1$\,dex in the HII model \citep{dagostino_comparison_2019} and AGN narrow line region model \citep{zhu_new_2023}. Among the two photoionization codes, only MAPPINGS has the ability to calculate shock models, {\color{black}including the postshock and photoionized preshock (precursor) emission.} We adopt the atomic data from the CHIANTI version 10 database \citep{del_zanna_chiantiatomic_2021} for the lightest 30 elements. 

The abundance scaling relations in all theoretical models are adopted from \citet{2017MNRAS.466.4403N}, which use the local B-stars abundance set from \citet{nieva_present-day_2012} and measured abundance scaling relations based on the observations of nearby HII regions \citep[e.g.][]{izotov_heavy-element_1999,israelian_galactic_2004,spite_first_2005,fabbian_co_2009,nieva_present-day_2012}. The main difference between local B-star abundance set and the solar photospheric abundance set in \citet{asplund_chemical_2009} is a 0.05\,dex enhancement in the oxygen abundance and a 0.03\,dex enhancement in the iron abundance \citep{sutherland_effects_2017}. In addition, the abundance scaling relations in \citet{2017MNRAS.466.4403N} are more realistic and sophisticated than the uniform scaling relations commonly applied. 

We consider dust in all theoretical models. We adopt the dust depletion factors from \citet{jenkins_unified_2009} where the depletion value for Fe is -1.5\,dex, a value derived from Wide Integral Field Spectrograph observations of local and Magellanic Cloud HII regions (Nicholls et al., in preparation). In shock models, dust destruction is also considered since observations show that dust is largely destroyed in the shock \citep{beck-winchatz_gas-phase_1996,reipurth_hubble_2000,dopita_spectral_1995}. 

\subsection{HII region Photoionization Model}

We adopt the isobaric HII region photoionization model from \citet{kewley_theoretical_2019}. The stellar ionizing radiation field is generated from Starburst99 by adopting the Salpeter initial mass function \citep{salpeter_luminosity_1955} with an upper mass limit of 100$M_\odot$, the Pauldrach/Hillier stellar atmosphere model \citep{hillier_treatment_1998,pauldrach_radiation-driven_2001}, and the Geneva group ``high'' mass-loss evolutionary tracks from \citet{meynet_grids_1994}. We obtain the stellar ionizing radiation field at the stellar age of 5\,Myr with a continuous star formation history for the HII region model.

In the HII region model, we vary the metallicity in the range of $7.3\leq$12+$\log(\rm O/H)\leq9.4$, ionization parameter in the range of $-3.8\leq\log(U)\leq-1.0$, and pressure in the range of 6.2$\lesssim\log{(P/k)}\lesssim$7.8 to represent the majority of nearby HII regions. 


\subsection{AGN Photoionization Model}

We adopt the isobaric radiation-pressure-dominated AGN narrow line regions model in \citet{zhu_new_2023}. This AGN model uses the AGN ionizing radiation field generated from OXAF \citep{thomas_physically_2016}, a simple version of AGN radiation model OPTXAGNF \citep{done_intrinsic_2012,jin_combined_2012} that model the continuum emission radiated from AGN thin accretion disk and a Comptonizing corona surrounding a central rotating black hole. AGNs that have $10^6\lesssim M_{\rm BH}/M_{\odot}\lesssim10^9$ and $0.05\lesssim L/L_{\rm Edd}\lesssim0.3$ have thin accretion disks \citep{novikov_astrophysics_1973,done_intrinsic_2012,thomas_physically_2016}. The most important parameter in the OXAF model is $E_{\rm peak}$, the peak energy of the accretion disk thermal emission, which can change the AGN optical emission line ratios by more than $\sim0.5$\,dex \citep{thomas_physically_2016,zhu_new_2023}. A representative range of $E_{\rm peak}$ for most AGNs in thin-disk accretion mode is $-2.0\leq\log E_{\text{peak}}/(\text{keV})\leq-1.0$.

In the AGN region model, we vary the metallicity in the range of $7.3\leq$12+$\log(\rm O/H)\leq9.4$, ionization parameter in the range of $-3.8\leq\log(U)\leq-1.0$, pressure in the range of 6.2$\lesssim\log{(P/k)}\lesssim$7.8, and $E_{\rm peak}$ in the range of $-2.0\leq\log E_{\text{peak}}/(\text{keV})\leq-1.0$ to represent the majority of nearby AGN narrow-line regions.

\subsection{Radiative Shock Model}

{\color{black}In this work, we consider two grids of shock models: a `pure shock' model grid and a `fast shock' model grid. The `pure shock' model grid is mainly contributed by post-shock emission and has little to no ($\leq10\%$, corresponding to $\lesssim$0.1 dex modeling uncertainty \citep{kewley_theoretical_2001}) precursor emission, while the `fast shock' model grid has a significant contribution from the photoionized precursor emission.}

We adopt the radiative shocks model in \citet{sutherland_effects_2017}. This radiative shock model applied the first fully self-consistent treatment of the preshock and postshock regions by using a multizone iterative scheme in solving the time-dependent photoionization, recombination, photoelectric heating, and line cooling processes throughout the preshock and postshock regions. The utility of this multizone iterative scheme overcomes the main shortcoming in previous shock models of using a single mean field on the final time-dependent ionization and temperature.

These shock models provide four classes of solutions for the preshock region (also called `precursor' in \citet{dopita_spectral_1996,allen_mappings_2008,sutherland_effects_2017}) over the shock velocity range $10\,\rm km\,s^{-1}\lesssim v_s\lesssim1500\,\rm km\,s^{-1}$, including a cold neutral precursor at $v_s\lesssim40\,\rm km\,s^{-1}$, warm neutral precursor at $40\,\rm km\,s^{-1}\lesssim v_s\lesssim75\,\rm km\,s^{-1}$, warm partly ionized precursor at $75\,\rm km\,s^{-1}\lesssim v_s\lesssim150\,\rm km\,s^{-1}$, and fully-ionized precursor at $v_s\gtrsim150\, \rm km\,s^{-1}$.

Given the nature of AGN-driven shocks, we only include the fast shock model ($v_s\gtrsim150\,\rm km\,s^{-1}$) in this work. In the fast shock regime, the postshock temperature ($T_s\propto v_s^2$ for strong shocks with $\cal{M}\gg\,$1, \citet{shull_theoretical_1979}) is high enough to generate a strong EUV photon field in the cooling zone that can fully ionize the preshock region and take it to photoionization equilibrium \citep{sutherland_cooling_1993,dopita_spectral_1995,dopita_spectral_1996}. This photoionization process in fast shock precursors is similar to a plane parallel HII region \citep{allen_mappings_2008}. The line emission from this fully ionized photoionization equilibrium fast shock precursor can significantly contribute to the total spectrum emitted by a fast shock \citep{dopita_spectral_1996,allen_mappings_2008}. 

In addition to shock velocity, other parameters in the shock model are the transverse component of the preshock magnetic field $B_0$, metallicity 12+$\log(\rm O/H)$, and gas pressure $P_0$ in the preshock gas. For convenience in the model calculation, a dimensionless magnetic-to-ram pressure ratio $\eta_M=B_0^2/(4\pi\rho_0 v_s^2)$ is defined and directly used in the shock model, where $\rho_0$ is the mass density of the preshock gas. We vary the shock velocity in the range of $150\,\rm km\,s^{-1}\leq v_s\leq1500\,\rm km\,s^{-1}$, the metallicity in the range of $7.3\leq$12+$\log(\rm O/H)\leq9.4$, gas pressure in the range of 6.2$\leq \log{(P/k)}\leq $10.2, and the magnetic-to-ram pressure ratio in the range of $0\leq \eta_M \leq 0.1$ (corresponding to the magnetic field strength in the range of $0\leq B_0\lesssim1.721\times10^3 \mu G$.) 

{\color{black} The shock models in this work are computed by summing the emission from the shocked and the precursor (photoionized preshock) regions, assuming a fixed ratio of H$\beta$ luminosities between the precursor and the shock, $L_{\mathrm{H}\beta,\mathrm{precursor}} : L_{\mathrm{H}\beta,\mathrm{shock}}$. This ratio varies in different shock regions, and is affected by the shock velocity and the extent of post-shock and precursor regions. In this work, we adjust the contribution from the shocked and the precursor to H$\beta$ luminosities in increments of 0.1 and select the best-fit shock model to match the data via visual inspection. We find a pure shock-dominated model with $L_{\mathrm{H}\beta,\mathrm{precursor}} : L_{\mathrm{H}\beta,\mathrm{shock}}=$0.1:0.9 and a precursor-dominated model with $L_{\mathrm{H}\beta,\mathrm{precursor}} : L_{\mathrm{H}\beta,\mathrm{shock}}=$0.7:0.3 provide the best-fit to two shock sequences of the NGC\,5728 MUSE IFU data. In the following sections, we use `pure shock' to refer to the pure shock-dominated models, and `fast shock' to refer to the precursor-dominated models.} 

\section{New 3D diagram}

\begin{figure*}[htb]
\epsscale{1.1}
\plotone{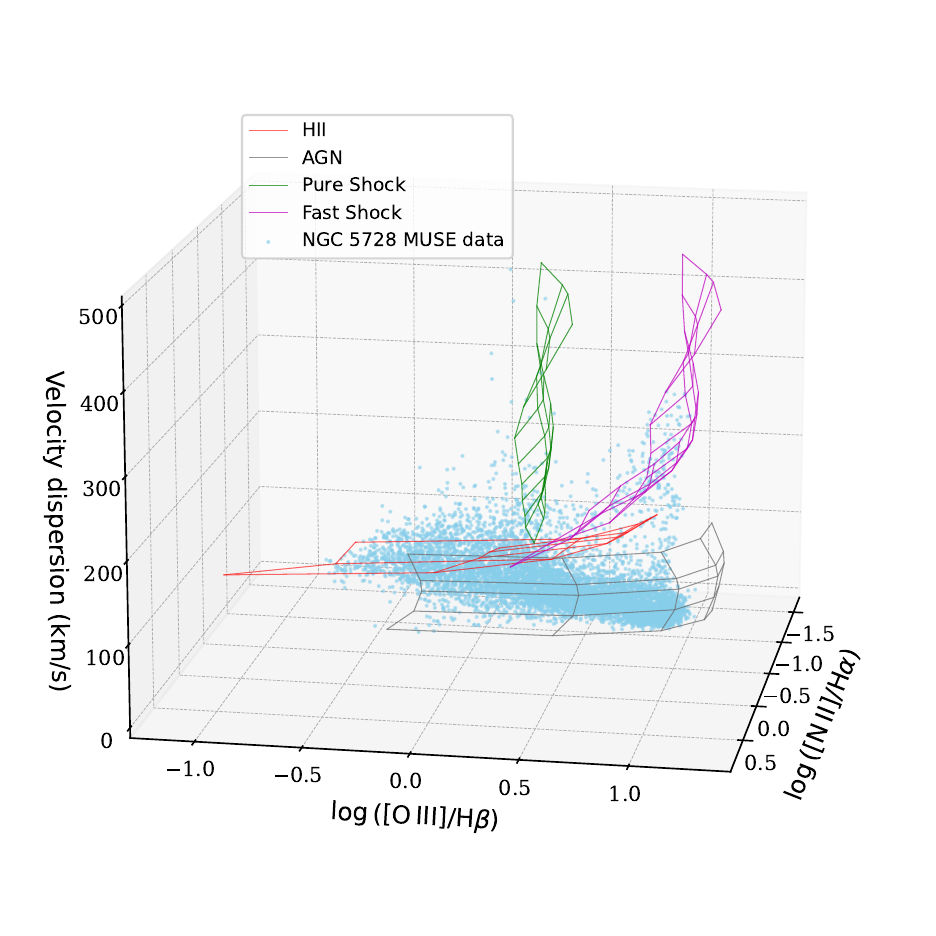}
\caption{The distribution of MUSE IFU data for NGC\,5728 on the new theoretical 3D diagram. The HII with $\log{P/k}=6.0$ and AGN models with $\log{P/k}=7.4$ and $\log E_{\text{peak}}/(\text{keV})=-1.0$ are shown on the diagram in red and black grids that consist of constant metallicity lines ($12+\log(\rm O/H)=8.12, 8.42, 8.82$ for HII model and $12+\log(\rm O/H)=8.43, 8.70, 8.80, 9.02, 9.26$ for AGN model) and constant ionization parameter lines ($\log(\rm U)=-3.75, -3.5, -3.25, -3.0, -2.5, -2.25$ for HII model and $\log(\rm U)=-3.8, -3.4, -3.0, -2.6, -2.2$ for AGN model). The pure shock and fast shock models with $\log{(P/k)}=10.2$ and $\eta_M=0.0001$ are shown on the diagram in green and magenta grids with constant metallicity lines ($12+\log(\rm O/H)=8.43, 8.70, 8.80, 9.02, 9.26$) and constant shock velocity lines ($V_s=158, 179, 202, 228, 257, 290, 328, 370, 418, 472$\,km/s). The MUSE IFU data for NGC\,5728 is shown with blue dots on the diagram. {\color{black}An interactive version of this figure is available in the online journal.} \label{fig:1}}
\end{figure*}

Utilizing the most up-to-date theoretical models for the HII region, AGN narrow line regions, and radiative shocks, we explore the theoretical 3D diagrams where the theoretical models for different mechanisms are well-separated. We find the 3D diagram that consists of emission line ratios [N~II]$\,\lambda6584$/H$\alpha$, [O~III]$\,\lambda5007$/H$\beta$, and emission line velocity dispersion presents a remarkable ability in distinguishing HII, AGN, and shock models. This 3D diagram is inspired by the successful separation of HII and AGN regions on the BPT diagram and the significant kinematic difference reflected in the [OIII] velocity dispersion ($\sigma_{\rm O[III]}$) in fast shock regions ($\sigma_{\rm O[III]}\sim300-1000$\,km/s, \citet{dagostino_separating_2019}), AGN regions ($\sigma_{\rm O[III]}\approx147$\,km/s, \citet{woo_prevalence_2016,joh_gas_2021}), and HII regions ($\sigma_{\rm O[III]}\approx58$\,km/s, \citet{joh_gas_2021}). {\color{black} This theoretical 3D diagram is relatively insensitive to the dust extinction in the observation, because [N~II]$\,\lambda6584$ and H$\alpha$, as well as [O~III]$\,\lambda5007$ and H$\beta$ are closely spaced in wavelength. Additionally, velocity dispersion measurement is not affected by dust extinction.}

From the theoretical models, we can obtain the theoretical values of emission line ratios [N~II]$\,\lambda6584$/H$\alpha$ and [O~III]$\,\lambda5007$/H$\beta$ in HII, AGN, and shock regions, respectively. However, theoretical models do not predict velocity dispersion because velocity dispersion is determined by gas kinematics, which is not considered in calculating HII and AGN photoionization models. For shock models, the shock velocity can serve as a proxy for the velocity dispersion in shock regions, assuming an upright viewing angle. {\color{black} Projection effects might result in the observed velocity dispersion being lower than the shock velocity. Therefore, this 3D diagram is most effective in identifying shocks that propagate toward the observer.}

A statistical study of the [O~III]$\,\lambda5007$ emission line velocity dispersion in Seyfert galaxies and star-forming galaxies in MaNGA survey shows that $\sigma_{\rm O[III]}$ varies across different galaxies \citep{joh_gas_2021}. Therefore, the velocity dispersion for HII and AGN models to be used in the 3D diagram should be determined for each galaxy empirically based on its observational data.

We estimate the average velocity dispersion from the HII-dominated and AGN-dominated spaxels in the NGC\,5728 IFU data for the HII and AGN models used in Figure~\ref{fig:1}. The HII-dominated spaxels are chosen based on the criterion that their [O~III]$\,\lambda5007$/H$\beta$ line ratio is lower than the star-formation and AGN separation line from \citet{kauffmann_host_2003} on the BPT diagram. The AGN-dominated spaxels are selected from the criterion that their [O~III]$\,\lambda5007$/H$\beta$ line ratio is higher than the star formation and AGN separation line from \citet{kewley_theoretical_2001} on the BPT diagram. {\color{black}In addition, we also apply an upper limit on the velocity dispersion in selecting HII-dominated spaxels ($\sigma_{\rm O [III]}\lesssim120$\,km/s) and AGN-dominated spaxels ($\sigma_{\rm O[III]}\lesssim400$\,km/s). These upper limits are chosen based on statistical studies of MaNGA galaxies \citep{joh_gas_2021,deconto-machado_ionised_2022,riffel_mapping_2023}. From the observational data, we obtain the average velocity dispersion $\overline{\sigma}_{\rm O[III]}=100$\,km/s for the HII-dominated spaxels and $\overline{\sigma}_{\rm O[III]}=125$\,km/s for the AGN-dominated spaxels in NGC5728. The HII model and AGN model are then fixed at their corresponding values on the velocity dispersion axis in the theoretical 3D diagram.

When multiple kinematic components are present in some spaxels of the IFU data, each component should be analyzed separately using the theoretical 3D diagram, because they might originate from distinct ionizing sources. In cases where multiple components are observed in only some (e.g., [O~III]) but not all (e.g., H$\beta$) emission lines, we recommend special treatment to interpret the kinematic complexity. The appropriate approach to incorporating such data into the theoretical 3D  diagram should be evaluated on a case-by-case basis, based on a careful assessment of the physical origin of the components.}

Figure~\ref{fig:1} shows an example of the theoretical 3D diagram and how the MUSE IFU data for NGC\,5728 is distributed on this 3D diagram. The HII and AGN models are shown on the diagram in red and black grids consisting of constant metallicity lines and ionization parameter lines. The pure shock and fast shock models are shown on the diagram in green and magenta grids with constant metallicity lines and constant shock velocity lines. The MUSE IFU data for NGC\,5728 is shown in blue on the diagram. 

In Figure~\ref{fig:1}, the NGC\,5728 spaxels form a low-velocity dispersion sequence that links the HII model region and AGN model region, indicating that both HII regions and AGN regions present within the field of view of the NGC\,5728 IFU data and some regions in this galaxy are excited by both AGN and HII region radiation. In addition, two vertical sequences are also revealed by the NGC\,5728 spaxels, one originating from the AGN model region toward the fast shock model region and the other linking the HII model region with the pure shock model region. The clear fast shock-AGN mixing sequences suggest the presence of fast shocks in this galaxy, and some of the spaxels are excited by fast shocks and AGN radiation at the same time. 

The coincidence between the theoretical models and observational IFU data on this 3D diagram enables the excitation sources in the galaxy IFU data to be unambiguously identified. In the following section, we use the 3D diagram to estimate the fractional contributions from all power sources in NGC\,5728. 

\section{Estimating the Fractional Contribution of Excitation Mechanisms}\label{sec:frac}

There is more than one approach to converting the information in the 3D diagram into the fractional contribution from each excitation mechanism. This section presents the approach that provides a reasonable separation, which agrees with multi-wavelength observations. 

To estimate the fractional contribution of HII, AGN, and shocks for each spaxel based on its location on the 3D diagram, we first select four basis spaxels that represent 100\% star-forming, 100\% AGN, 100\% fast shock, and 100\% pure shock in this 3D diagram. The 100\% star-forming spaxel is the spaxel that has the lowest [O~III]$\,\lambda5007$/H$\beta$ ratio from the HII-dominated spaxels defined above. The 100\% AGN spaxel is the spaxel that has the highest [O~III]$\,\lambda5007$/H$\beta$ ratio from the AGN-dominated spaxels defined above. We chose the 100\% fast shock spaxel from the maximum point where the AGN-fast shock mixing sequence intersects with the fast shock model grid. Similarly, the 100\% pure shock spaxel is chosen to be the maximum point where the HII-pure shock mixing sequence intersects with the pure shock model grid. The determination of 100\% shock spaxels involves visual inspection. 

After selecting four basis spaxels, we calculate the fractional contribution of each mechanism for each spaxel in the following steps. In the following equation, $x=\log$([N~II]/H$\alpha$), $y=\log$([O~III]/H$\beta$), and $z=\,\sigma_{\rm O[III]}\,$(km/s).

Step(i). Calculate the minimal ($x_{min}$,$y_{min}$,$z_{min}$) and maximal($x_{max}$,$y_{max}$,$z_{max}$) coordinates of basis spaxels. In NGC\,5728, four basis spaxels 100\% star-forming spaxel ($x_1$,$y_1$,$z_1$), 100\% AGN spaxel ($x_2$,$y_2$,$z_2$), 100\% fast shock spaxel ($x_3$,$y_3$,$z_3$), and 100\% pure shock spaxel ($x_4$,$y_4$,$z_4$) are selected. Therefore, the minimal and maximal coordinates can be obtained through:
\begin{align*}
    x_{min} &= \min(x_1,x_2,x_3,x_4), \quad x_{max}&=\max(x_1,x_2,x_3,x_4), \\
    y_{min} &= \min(y_1,y_2,y_3,y_4), \quad y_{max}&=\max(y_1,y_2,y_3,y_4), \\
    z_{min} &= \min(z_1,z_2,z_3,z_4), \quad z_{max}&=\max(z_1,z_2,z_3,z_4),
\end{align*}

Step(ii). Calculate the normalized distance $d_{n,{\rm HII}}$, $d_{n,{\rm AGN}}$, $d_{n,{\rm fs}}$, and $d_{n,{\rm ps}}$ from each spaxel ($x_i$,$y_i$,$z_i$) to the 100\% star-forming spaxel, 100\% AGN spaxel, 100\% fast shock spaxel, and 100\% pure shock spaxel. The normalization factor is the difference between the minimal and maximal coordinates calculated above. The normalized distance in x and y axes are firsted converted from log scale ($d_{log}$) to linear scale ($d_{linear}$) via $d_{linear}=10^{d_{log}}/(10-1)$. Then the normalized distance is calculated through:
\begin{scriptsize}
\begin{align*}
    d_{n,{\rm HII}} &= \sqrt{\left(\frac{10^{\frac{x_i - x_1}{x_{\max} - x_{\min}}}}{9}\right)^2 + \left(\frac{10^{\frac{y_i - y_1}{y_{\max} - y_{\min}}}}{9}\right)^2 + \left(\frac{z_i - z_1}{z_{\max} - z_{\min}}\right)^2} \\
    d_{n,{\rm AGN}} &= \sqrt{\left(\frac{10^{\frac{x_i - x_2}{x_{\max} - x_{\min}}}}{9}\right)^2 + \left(\frac{10^{\frac{y_i - y_2}{y_{\max} - y_{\min}}}}{9}\right)^2 + \left(\frac{z_i - z_2}{z_{\max} - z_{\min}}\right)^2} \\
    d_{n,{\rm fs}} &= \sqrt{\left(\frac{10^{\frac{x_i - x_3}{x_{\max} - x_{\min}}}}{9}\right)^2 + \left(\frac{10^{\frac{y_i - y_3}{y_{\max} - y_{\min}}}}{9}\right)^2 + \left(\frac{z_i - z_3}{z_{\max} - z_{\min}}\right)^2} \\
    d_{n,{\rm ps}} &= \sqrt{\left(\frac{10^{\frac{x_i - x_4}{x_{\max} - x_{\min}}}}{9}\right)^2 + \left(\frac{10^{\frac{y_i - y_4}{y_{\max} - y_{\min}}}}{9}\right)^2 + \left(\frac{z_i - z_4}{z_{\max} - z_{\min}}\right)^2}
\end{align*}
\end{scriptsize}

Step(iii). Calculate the fractional contribution based on the normalized distances. The closer a spaxel is to a basis spaxel, the larger the contribution from the mechanism represented by the basis spaxel. We can therefore convert the relative inverse distance into the fractional contribution values:
\begin{align*}
	f_{\rm HII}= \frac{1/d_{n,{\rm HII}}}{1/d_{n,{\rm HII}}+1/d_{n,{\rm AGN}}+1/d_{n,{\rm fs}}+1/d_{n,{\rm ps}}} \\
	f_{\rm AGN}= \frac{1/d_{n,{\rm AGN}}}{1/d_{n,{\rm HII}}+1/d_{n,{\rm AGN}}+1/d_{n,{\rm fs}}+1/d_{n,{\rm ps}}} \\
	f_{\rm fs}= \frac{1/d_{n,{\rm fs}}}{1/d_{n,{\rm HII}}+1/d_{n,{\rm AGN}}+1/d_{n,{\rm fs}}+1/d_{n,{\rm ps}}} \\
	f_{\rm ps}= \frac{1/d_{n,{\rm ps}}}{1/d_{n,{\rm HII}}+1/d_{n,{\rm AGN}}+1/d_{n,{\rm fs}}+1/d_{n,{\rm ps}}}
\end{align*}

Step(iv). To better present the 100\% HII-dominated, AGN-dominated, fast shock-dominated, and pure shock-dominated regions in the IFU data, we define a neighbor area $d_n < 0.2$ for each basis spaxel and assign a 100\% contribution from the mechanism to all spaxels located in this neighbor area:
\begin{align*}
	\textbf{If}\quad d_{n,{\rm HII}}<0.2, f_{\rm HII}=1.0, f_{\rm AGN},f_{\rm fs},f_{\rm ps}=0 \\
	\textbf{If}\quad d_{n,{\rm AGN}}<0.2, f_{\rm AGN}=1.0, f_{\rm HII},f_{\rm fs},f_{\rm ps}=0 \\
	\textbf{If}\quad d_{n,{\rm fs}}<0.2, f_{\rm fs}=1.0, f_{\rm HII},f_{\rm AGN},f_{\rm ps}=0 \\
	\textbf{If}\quad d_{n,{\rm ps}}<0.2, f_{\rm ps}=1.0, f_{\rm HII},f_{\rm AGN},f_{\rm fs}=0 
\end{align*}

Step(v). Assign 100\% shock contribution to spaxels lying above the 100\% shock basis spaxel. Recall that the 100\% fast shock basis spaxel is located where the AGN-shock mixing sequence intersects with the fast shock model, {\color{black} and the 100\% pure shock basis spaxel is located where the HII-shock mixing sequence intersects with the pure shock model.} Spaxels above these 100\% shock basis spaxel should all be shock-dominated since their distances to the 100\% HII basis spaxel and 100\% AGN basis spaxel are further than the 100\% shock basis spaxel. Therefore, we need to adjust the fractional contribution for spaxels that lie above the 100\% fast and pure shock basis spaxels. Addtionally, we use $\log(\rm [O~III]/H\beta)=0.7$ to discriminate between the 100\% fast shock and 100\% pure shock spaxels:
\begin{align*}
	\textbf{If}\quad z_i>z_3 \,\,\textbf{and}\,\, y_i>0.7, f_{\rm fs}=1.0, f_{\rm HII},f_{\rm AGN},f_{\rm ps}=0 \\
	\textbf{If}\quad z_i>z_4 \,\,\textbf{and}\,\, y_i<=0.7, f_{\rm ps}=1.0, f_{\rm HII},f_{\rm AGN},f_{\rm fs}=0
\end{align*}

{\color{black}Step(vi). To estimate the uncertainties in the calculated fractional contributions, we performed a Monte Carlo error propagation analysis. We first obtain the observational uncertainties for $\sigma_{\rm O[III]}\,$(km/s) as $z_{err}$, and calculate the uncertainties for $x_{err}$ and $y_{err}$ using the observational uncertainties in corresponding emission line fluxes. The logarithmic line ratio errors were calculated using standard error propagation:
{\small
\begin{equation*}
\sigma_{\log(X/Y)} = \frac{1}{\ln 10} \sqrt{ \left( \frac{\sigma_X}{X} \right)^2 + \left( \frac{\sigma_Y}{Y} \right)^2 },
\end{equation*}}where $X$ and $Y$ are the line fluxes and $\sigma_X$, $\sigma_Y$ are their uncertainties. We then generated $N = 200$ realizations ($x'$,$y'$,$z'$) of each spaxel by sampling from Gaussian distributions centered on the observed values ($x,y,z$), with standard deviations equal to the observational uncertainties ($x_{err},y_{err},z_{err}$). For each realization, we then recalculate the fractional contributions $f_{\rm HII}'$,$f_{\rm AGN}'$,$f_{\rm fs}'$,$f_{\rm ps}'$) by repeating Steps (ii) to (v). The final contribution uncertainties were taken as the standard deviation of the contribution values across all realizations:
{\small
\begin{equation*}
\sigma_f = \sqrt{ \frac{1}{N - 1} \sum_{j=1}^{N} \left( f'_j - \bar{f'} \right)^2 },
\end{equation*}}where $f'_j$ is the fractional contribution from a single excitation source for a single realization and $\bar{f'}$ is the mean contribution of $f'_j$ over $N$ realizations. 

This approach captures the propagation of observational uncertainties into the calculation of the fractional contribution, and provides us with an evaluation of how much the calculation depends on measurement uncertainties and is affected by the model degeneracies in the certain regime of the theoretical 3D diagram.
}

The above steps are designed for galaxies that present HII, AGN, fast shock, and pure shock mixing on the 3D diagram. For galaxies with fewer mechanisms presented in the 3D diagram, fractional contributions from each mechanism can be calculated by adjusting the number of basis spaxels in the above calculation. 

\section{Excitation Mechanisms of NGC\,5728}

\begin{figure*}[htb]
\epsscale{0.54}
\plotone{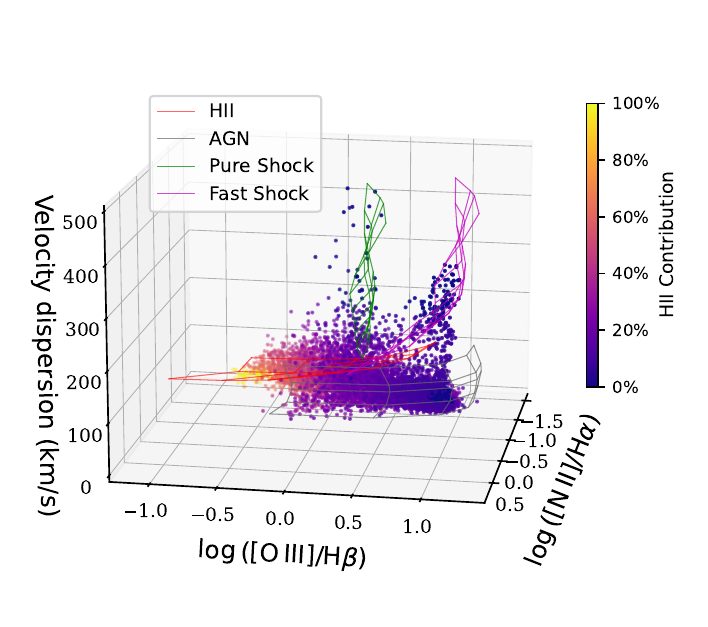}
\plotone{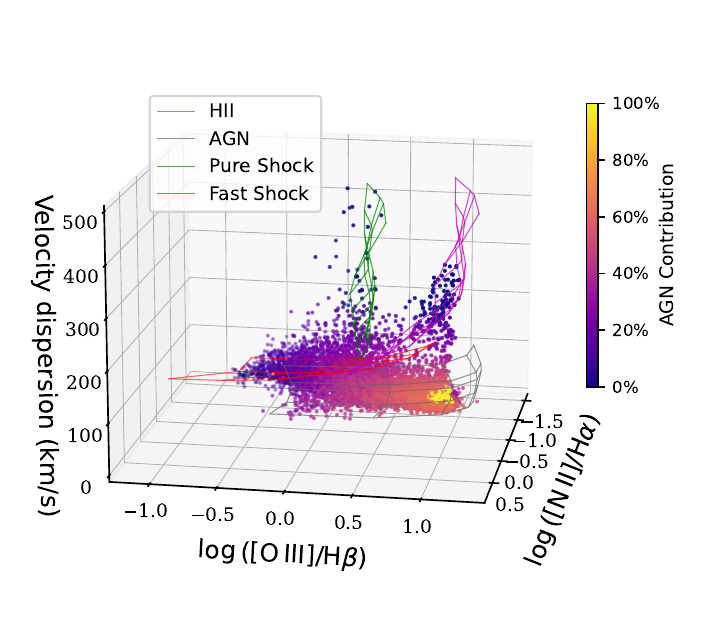}
\plotone{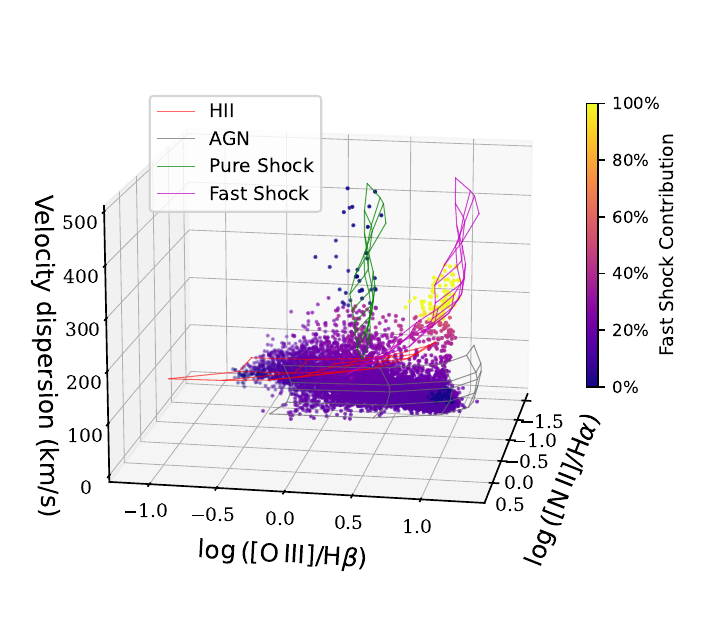}
\plotone{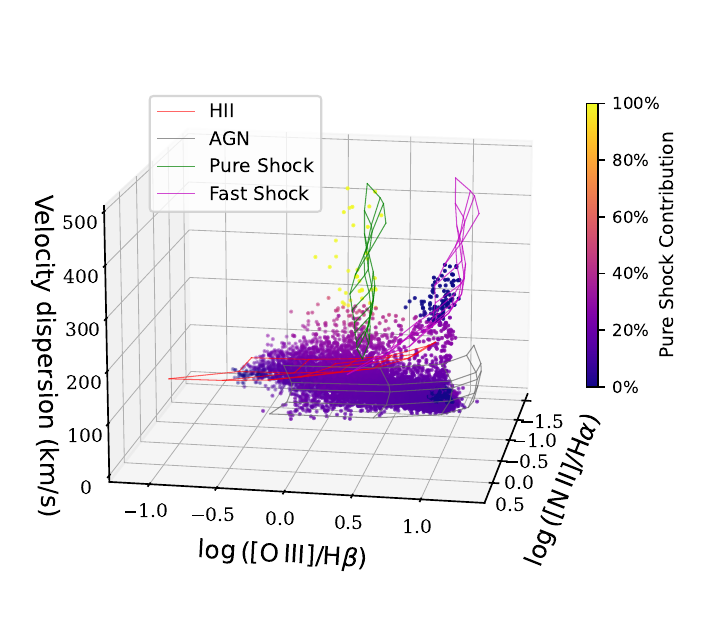}
\caption{The distribution of MUSE IFU data for NGC\,5728 on the new theoretical 3D diagram, with the spaxels color-coded by HII contribution fraction (top left), AGN contribution fraction (top right), fast shock contribution fraction (bottom left), and pure shock contribution fraction (bottom right). {\color{black}An interactive version of this figure is available in the online journal.}\label{fig:2}}
\end{figure*} 

\begin{figure*}[htb]
\epsscale{1.2}
\plotone{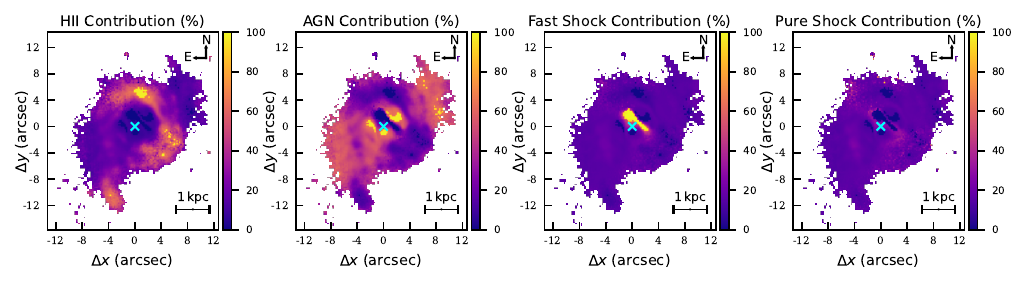}
\plotone{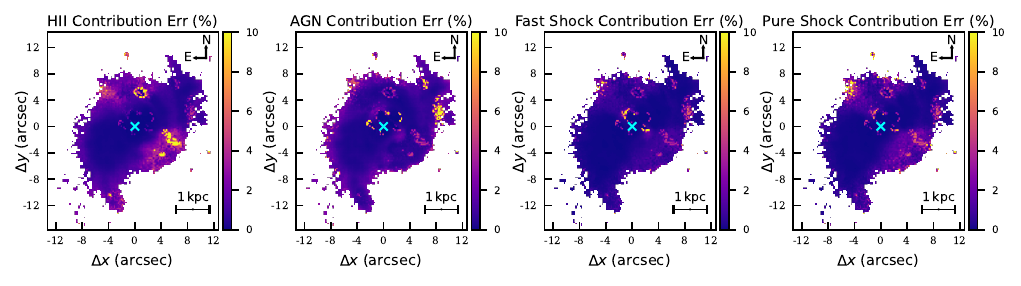}
\caption{2D maps of the MUSE IFU data for NGC\,5728, with spaxels color-coded by fractional contribution in the upper panel {\color{black}and the uncertainty in the bottom panel.} From left to right, the distributions of HII contribution, AGN contribution, fast shock contribution, and pure shock contribution are shown, respectively. The spaxel with the highest [O~III] luminosity, indicating the galaxy center, is marked with a cyan cross.\label{fig:4-21}}
\end{figure*}

We apply the MUSE IFU data for NGC\,5728 to the 3D diagram and find the mixing of HII, AGN, fast shocks, and pure shocks. Following the steps in section~\ref{sec:frac}, we separate the HII, AGN, and shock in NGC\,5728 and present the distribution of the fractional contribution on the 3D diagram in Figure~\ref{fig:2} and the galaxy two-dimensional (2D) map in Figure~\ref{fig:4-21}. 

As shown in Figure~\ref{fig:2}, spaxels that have $>$80\% contribution from HII, AGN, fast shocks, and pure shocks are distributed closely around the HII model, AGN model, fast shock model, and pure shock models, respectively. In contrast, spaxels have $\sim40-60$\% contribution generally located in the middle of the two models. This result suggests that the method in section~\ref{sec:frac} provides a reasonable conversion from space coordinates to the fractional contribution of each excitation mechanism in the 3D diagram.

Figure~\ref{fig:4-21} shows how HII, AGN, fast shocks, and pure shocks are distributed in the 2D map of NGC\,5728. In the upper left panel, the HII contribution map reveals that star formation mainly occurs in a ring structure surrounding the galaxy center with a projected radius of $\sim1\,$kpc in the sky plane, which was first noticed by \citet{schommer_ionized_1988}, and later studied by \citet{durre_agn_2018}, who used the BPT diagram to separate HII and AGN regions in NGC\,5728 with MUSE IFU data and reveal a star-forming ring structure. Following that, \citet{shin_positive_2019} found a coincided ring structure in the CO (2-1) emission from ALMA data, suggesting the ring structure is a gas-rich star-forming region.

In the upper left-middle panel, the AGN contribution map shows that AGN excitation dominates a bicone structure with a projected length of $\sim5\,$kpc in the sky plane from the southeast to the northwest direction, which is also suggested by \citet{shin_positive_2019,durre_agn_2018,shimizu_multiphase_2019} based on HII-AGN separation through the BPT diagram with MUSE IFU data and confirmed with the Chandra X-ray data tracing the ionization cones. 

In the upper right-middle panel, the fast shock contribution map reveals the presence of fast shocks in the galaxy center with a rectangular shape perpendicular to the AGN bicone structure with a projected length of $\sim1\,$kpc in the sky plane. This is the first detection of fast shocks in the center of NGC\,5728. The presence of shocks in molecular gas is also reported by \citet{davies_gatos_2024}, who used JWST/MIRI MRS data to study the mid-infrared rotational H$_2$ lines. In the nucleus region where we detect the fast shock, they found the H$_2$ line ratios characterize a power-law temperature distribution that can be explained by slow-shock excitation with shock velocity $V_s\approx 30$\,km/s.

In the upper right panel, the pure shock contribution map reveals a layer of shock emission encircling the central fast-shock regions. Additionally, pure shock emission extends outward to approximately $\sim2\,$kpc from the galaxy center along the cross-cone direction. Spaxels with 100\% pure shock emission are sparsely distributed near the edges of the galaxy 2D map. Given that all spaxels shown in the map have S/N > 5, and considering these 100\% pure shock spaxels are situated close to HII-dominated regions, they are likely genuine detections tracing shock activity near star-forming regions.

{\color{black}The bottom panel shows that the uncertainties in the estimated fractional contributions are generally within $\sim$5\%, with the dominant source of error arising from uncertainties in the velocity dispersion measurements. Regions with slightly higher uncertainties, reaching up to $\sim$10\%, are primarily associated with spaxels that were manually assigned a 100\% contribution based on their proximity to the basis spaxels in Step (iv). When Step (iv) is omitted from the calculation, the uncertainties in these regions drop below $\sim$2\%. Overall, the spatial distribution of uncertainties confirms that our contribution estimates are robust against both observational errors and model degeneracies.
}

\begin{figure*}[htb]
\epsscale{1.2}
\plotone{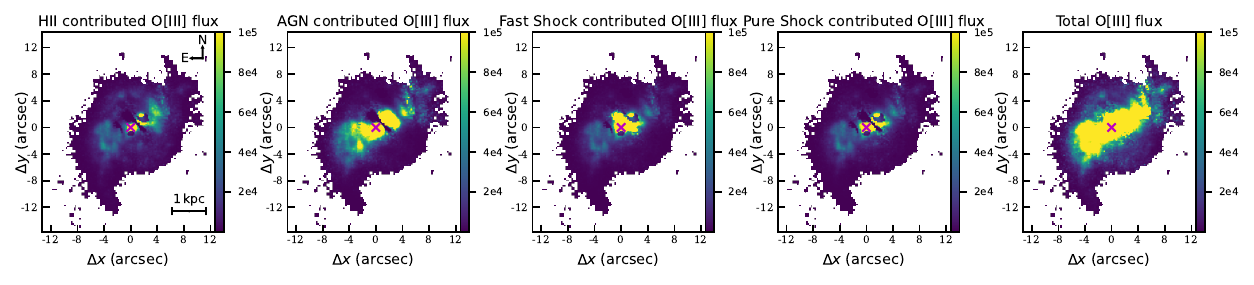}
\plotone{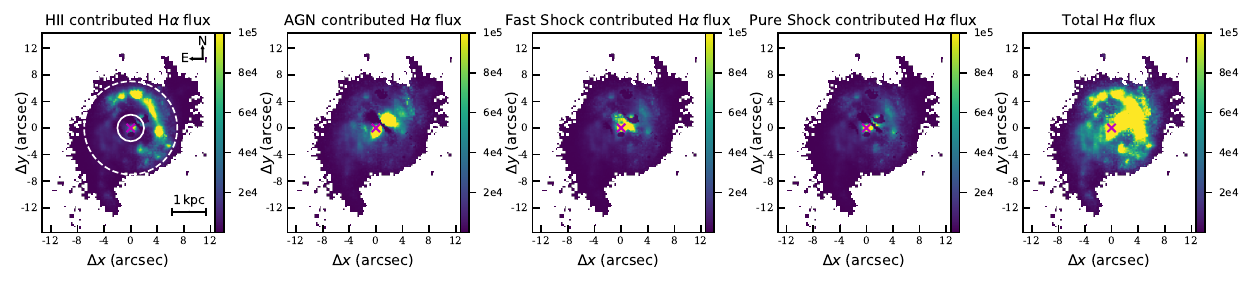}
\plotone{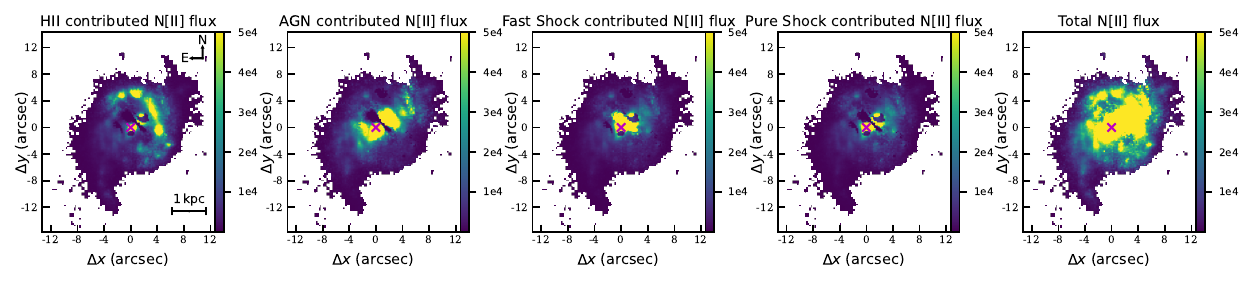}
\caption{2D maps of [O~III]$\,\lambda5007$ (the top panel), H$\alpha$ (the middle panel), and [N~II]$\,\lambda6584$ (the bottom panel) emission line fluxes of the MUSE IFU data for NGC\,5728. From left to right, each panel contains the emission line flux maps contributed by HII, AGN, fast shocks, pure shocks, and the sum of all four mechanisms. The unit of fluxes in this figure is $10^{-20}\,\rm erg\cdot s^{-1} \cdot cm^{-2}$. {\color{black}The spaxel with the highest [O~III] luminosity, indicating the galaxy center, is marked with a magenta cross. The solid and dashed white circle in the HII contributed H$\alpha$ flux map indicates the nuclear region with a radius of 2\arcsec and 6\arcsec, respectively.}
\label{fig:3}}
\end{figure*}

With the fractional contribution from each mechanism, we can isolate the emission line fluxes ($F_i$) contributed by HII ($F_{i,\rm HII}$), AGN ($F_{i,\rm AGN}$), fast shocks ($F_{i,\rm fast\,shock}$), and pure shocks ($F_{i,\rm pure\,shock}$) for each spaxel through:
\begin{align}
	F_{i,\rm HII} &= f_{\rm HII} \cdot F_i \\
    F_{i,\rm AGN} &= f_{\rm AGN} \cdot F_i \\
    F_{i,\rm fast\,shock} &= f_{\rm fs} \cdot F_i \\
    F_{i,\rm pure\,shock} &= f_{\rm ps} \cdot F_i
\end{align}

Figure~\ref{fig:3} presents the distribution of HII-contributed, AGN-contributed, fast shock-contributed, and pure shock-contributed emission line flux maps for [O~III]$\,\lambda5007$, H$\alpha$, and [N~II]$\,\lambda6584$ in NGC\,5728. {\color{black}All emission line fluxes are corrected for dust extinction following the steps in \citet{vogt_galaxy_2013} using $R_V^A=4.5$.} As a comparison, the right panel shows the total emission line maps before the separation. The top panel shows that the [O~III]$\,\lambda5007$ emission line is mainly present in the AGN-dominated bicone region. In the middle panel, H$\alpha$ emission is contributed by HII in the star-forming ring and by AGN and shock in the galaxy nuclear center. The [N~II]$\,\lambda6584$ emission line has a similar distribution as H$\alpha$ emission line but is dimmer at the HII-dominated ring and brighter at the AGN-dominated bicone and fast shock-dominated center region, consistent with the [N~II] line being more excited by the hard radiation field associated with AGN and shocks.

\begin{figure*}[htb]
\epsscale{0.55}
\plotone{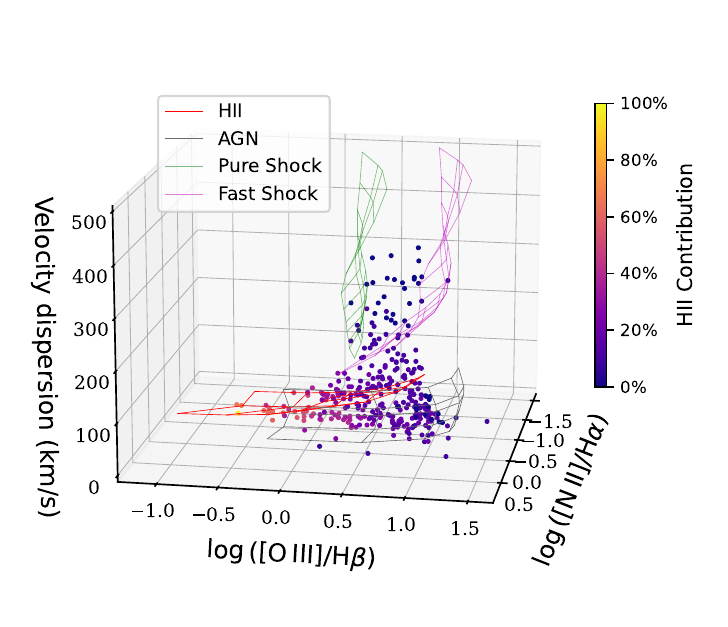}
\plotone{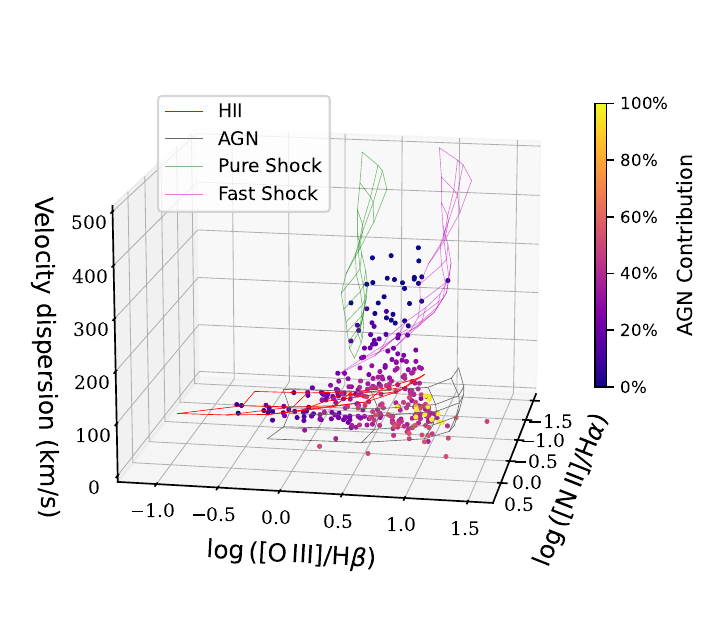}
\plotone{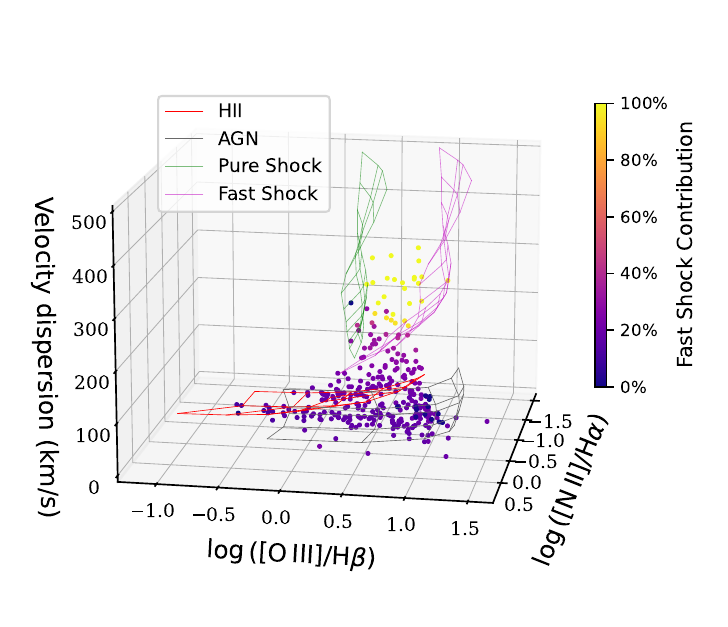}
\plotone{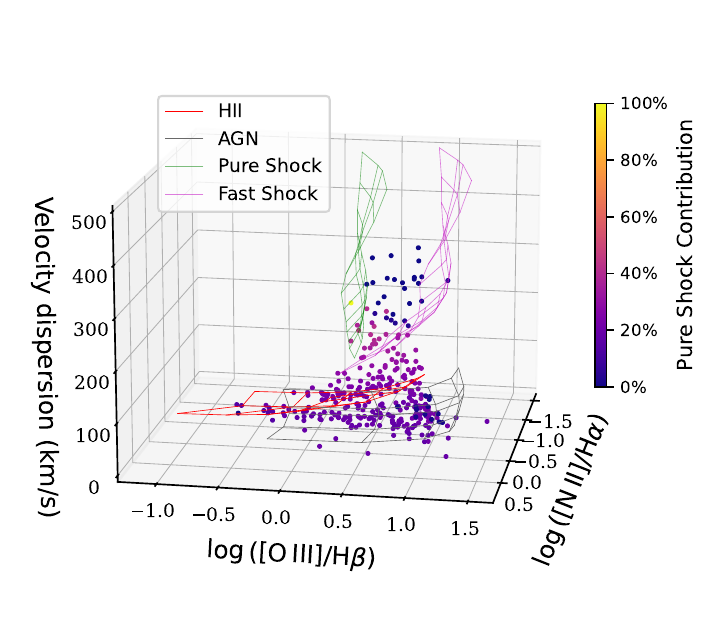}
\caption{The distribution of WiFeS IFU data for NGC\,5728 (observed in the S7 survey) on the new theoretical 3D diagram, with the spaxels color-coded by HII contribution fraction (top left), AGN contribution fraction (top right), fast shock contribution fraction (bottom left), and pure shock contribution fraction (bottom right). \label{fig:4}}
\end{figure*}

To test whether our 3D diagram is effective at lower spatial resolution, we compare our excitation mechanism separation with S7 IFU data for NGC\,5728 in Figure~\ref{fig:4}. Although the S7 IFU has a smaller field of view than the MUSE IFU, it is sufficient to include the central AGN-bicone region and the star-forming ring in NGC\,5728. Although having fewer spaxels and a larger spaxel size, the S7 IFU data also reveal the presence of HII, AGN, and shocks in the 3D diagram, as shown in figure~\ref{fig:4}.

\begin{figure*}[htb]
\epsscale{1.0}
\plotone{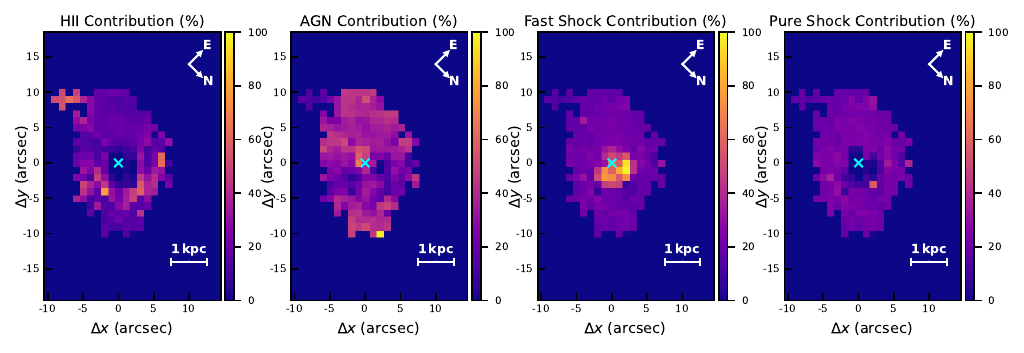}
\plotone{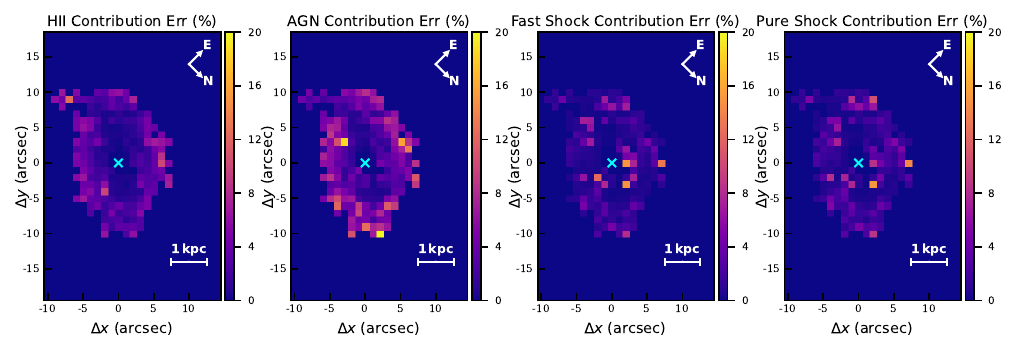}
\caption{2D maps of the S7 IFU data for NGC\,5728, with spaxels color-coded by fractional contribution in the upper panel {\color{black}and the uncertainty in the bottom panel.} From left to right, the distributions of HII contribution, AGN contribution, fast shock contribution, and pure shock contribution are shown, respectively. WiFes IFU was rotated by 135$^\circ$ in the anti-clockwise direction in the observation of NGC\,5728. {\color{black}The white arrows on the upper right corner denote the north and east in the sky. The spaxel with the highest [O~III] luminosity, indicating the galaxy center, is marked with a cyan cross. } \label{fig:4-22}}
\end{figure*}

Despite the lower spatial resolution of the S7 WiFeS IFU data (1$\arcsec$ or 200\,pc pixel$^{-1}$, compared to 0.2$\arcsec$ or 40\,pc pixel$^{-1}$ in MUSE), the distributions of HII-dominated, AGN-dominated, fast shock-dominated, and pure shock-dominated spaxels in Figure~\ref{fig:4-22} have similar shapes as in the MUSE IFU data in Figure~\ref{fig:4-21}. The HII-dominated spaxels are located along an off-center arc that is $\sim1\,$kpc away from the galaxy center in the northwest direction in the sky plane. This HII-dominated arc structure corresponds to the brightest section of the HII-dominated ring revealed by MUSE data. The AGN-dominated spaxels form a bicone structure with a projected length of $\sim5\,$kpc in the sky plane from southeast to northwest, as revealed by MUSE IFU data. The fast shock-dominated spaxels concentrate at the base of the AGN bicone structure, presenting a point-like region with a projected radius of $\sim1\,$kpc. This fast shock-dominated region corresponds to the center $\sim1\,$kpc of the rectangular fast shock-dominated region in MUSE IFU data. The point-like structure in S7 IFU data is likely a result of the broader seeing (1.2$\arcsec$, compared to 0.66$\arcsec$ in MUSE) and larger spaxel size (1$\arcsec$/pixel) compared to the MUSE IFU data (0.2$\arcsec$/pixel), which leads to a smoother appearance. The pure shock-dominated spaxels are $\sim1-2\,$kpc away from the galaxy center in the northeast and northwest direction, which is consistent with the MUSE IFU data. 

{\color{black}As shown in the bottom panel of Figure~\ref{fig:4-22}, the uncertainties in the fractional contributions derived from the S7 data are generally within $\sim$15\%, which is larger than the uncertainties obtained in the MUSE data. This increased uncertainty is primarily due to the higher noise level in the S7 observations, which have a shorter exposure time of $t = 1800 s$ compared to the deeper MUSE observations $ t = 4740 s$.}

\begin{figure*}[htb]
\epsscale{1.1}
\plotone{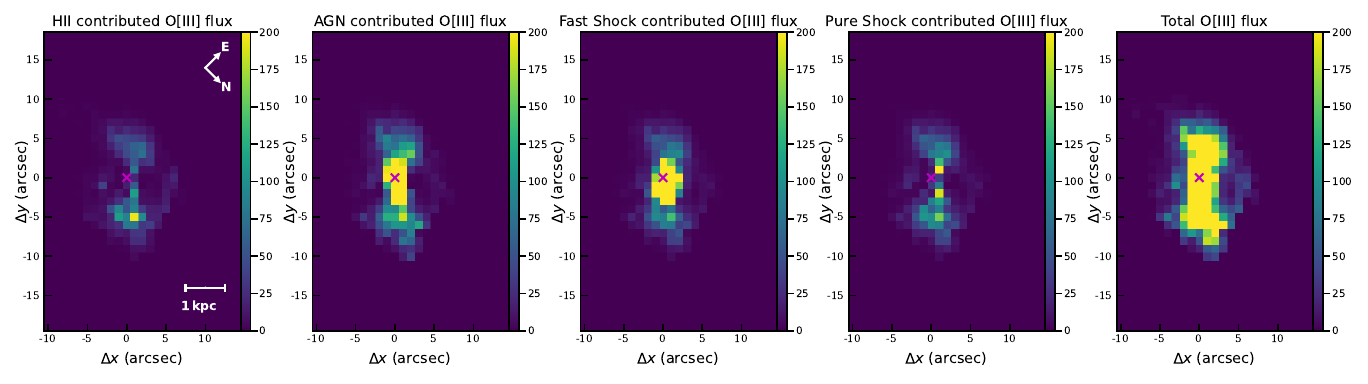}
\plotone{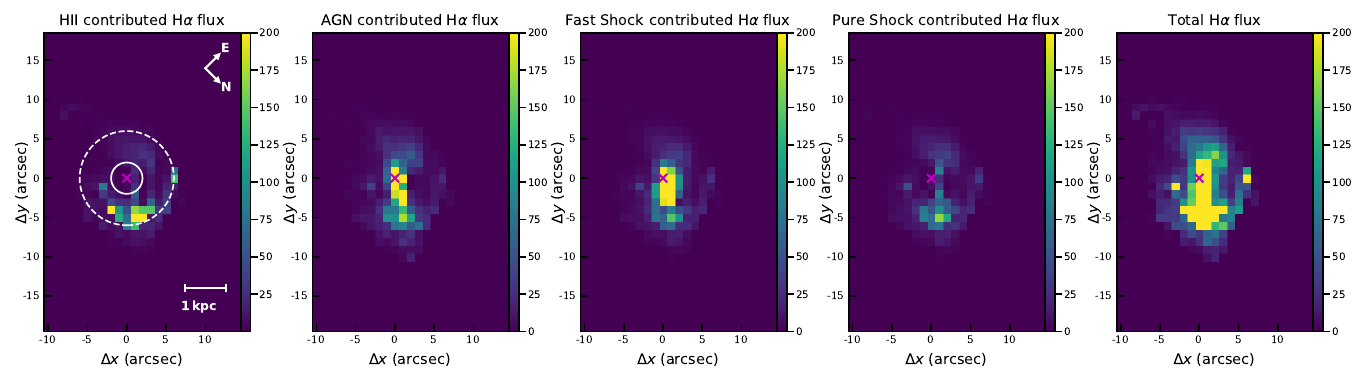}
\plotone{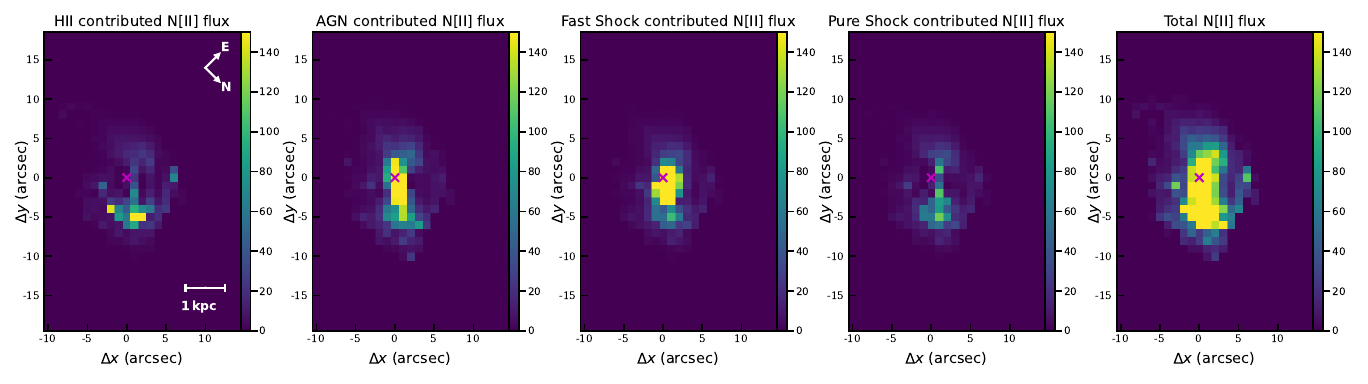}
\caption{2D maps of [O~III]$\,\lambda5007$ (the top panel), H$\alpha$ (the middle panel), and [N~II]$\,\lambda6584$ (the bottom panel) emission line fluxes of the MUSE IFU data for NGC\,5728. From left to right, each panel contains the emission line flux maps contributed by HII, AGN, fast shocks, pure shocks, and the sum of four mechanisms in total. The unit of fluxes in this figure is $10^{-16}\,\rm erg\cdot s^{-1} \cdot cm^{-2}$. The WiFes IFU was rotated by 135$^\circ$ in the anti-clockwise direction in the observation of NGC\,5728. {\color{black}The white arrows on the upper right corner denote the north and east in the sky. The spaxel with the highest [O~III] luminosity, indicating the galaxy center, is marked with a magenta cross. The solid and dashed white circle in the HII contributed H$\alpha$ flux map indicates the nuclear region with a radius of 2\arcsec and 6\arcsec, respectively.} \label{fig:5}}
\end{figure*}

Figure~\ref{fig:5} shows the 2D distribution maps of HII-contributed, AGN-contributed, fast shock-contributed, and pure shock-contributed [O~III]$\,\lambda5007$, H$\alpha$, and [N~II]$\,\lambda6584$ emission lines fluxes of the S7 WiFeS IFU data for NGC\,5728.

We also estimate the total fractional contribution from each mechanism ($f_{tot, \rm mec, EL}$) to the total flux ($F_{tot, \rm EL,corr}$) of emission lines [N~II]$\,\lambda6584$, H$\alpha$, [O~III]$\,\lambda5007$, H$\beta$, and [S~II]$\,\lambda\lambda$6717,31 in NGC\,5728. These total fractional contributions are calculated through the following equation:
\begin{equation}
	f_{tot,\rm mec,EL} =\sum (f_{i,\rm mec}\times F_{i,\rm EL})/F_{tot, EL}
\end{equation}
where $F_{i,\rm EL}$ is the extinction-corrected flux of an emission line in each spaxel. We calculate the total fractional contributions for all four mechanisms in MUSE IFU data and S7 WiFeS IFU data for spaxels with S/N$\geq$3 for emission lines [N~II], H$\alpha$, [O~III], H$\beta$. The results are shown in Table~\ref{tab:1}.

\setlength{\tabcolsep}{2.0pt}
\begin{deluxetable}{c|c|c|c|c|c}[hbt]
\small
\centering
\tablewidth{0pc}
\tablecaption{Total Fractional Contribution from HII, AGN, Fast shocks, and Pure shocks for Emission Lines \label{tab:1}}
\tablenum{1}
\tablehead{
\colhead{Emission Line} &
\colhead{$f_{\rm tot,\,HII}$} &
\colhead{$f_{\rm tot,\,AGN}$} &
\colhead{$f_{\rm tot,\,fs}$} &
\colhead{$f_{\rm tot,\,ps}$} &
\colhead{$F_{\rm tot}$} \\
\colhead{} &
\colhead{(\%)} &
\colhead{(\%)} &
\colhead{(\%)} &
\colhead{(\%)} &
\colhead{($10^{-13}$)}	
}
\startdata
\multicolumn{6}{c}{MUSE IFU data, S/N$\geq$3 for [N~II], H$\alpha$, [O~III], and H$\beta$}\\
\hline
{[O~III]$\,\lambda5007$} 			& 13.5$\pm$0.3 & 44.9$\pm$0.6 & 26.1$\pm$0.3 & 15.5$\pm$0.3 & 51.5$\pm$0.3 \\
{[N~II]$\,\lambda6584$} 			& 21.2$\pm$0.8 & 40.0$\pm$0.7 & 23.2$\pm$0.5 & 15.6$\pm$0.5 & 24.5$\pm$0.1 \\
{[S~II]$\,\lambda\lambda$6717,31} 	& 22.8$\pm$0.9 & 39.2$\pm$0.7 & 22.3$\pm$0.5 & 15.7$\pm$0.6 & 12.8$\pm$0.2 \\
H$\alpha$ 							& 29.5$\pm$1.2 & 34.3$\pm$0.7 & 20.1$\pm$0.5 & 16.1$\pm$0.7 & 28.3$\pm$0.2 \\
H$\beta$ 							& 29.5$\pm$1.2 & 34.3$\pm$0.7 & 20.1$\pm$0.5 & 16.1$\pm$0.7 & 9.9$\pm$0.3 \\
\hline
\multicolumn{6}{c}{S7 WiFeS IFU data, S/N$\geq$3 for [N~II], H$\alpha$, [O~III], and H$\beta$}\\
\hline
{[O~III]$\,\lambda5007$} 			& 8.6$\pm$0.8 & 37.6$\pm$1.8 & 42.6$\pm$0.9 & 11.2$\pm$0.9 & 39.6$\pm$1.4 \\
{[N~II]$\,\lambda6584$} 			& 13.0$\pm$1.1 & 34.2$\pm$1.8 & 40.3$\pm$1.0 & 12.5$\pm$1.0 & 19.4$\pm$0.7 \\
{[S~II]$\,\lambda\lambda$6717,31} 	& 13.2$\pm$1.3 & 35.2$\pm$2.0 & 38.6$\pm$1.0 & 13.0$\pm$1.0 & 9.8$\pm$0.6 \\
H$\alpha$ 							& 18.8$\pm$1.5 & 29.0$\pm$1.9 & 38.1$\pm$1.0 & 14.1$\pm$1.0 & 19.9$\pm$0.8 \\
H$\beta$ 							& 18.8$\pm$1.5 & 29.0$\pm$1.9 & 38.1$\pm$1.0 & 14.1$\pm$1.0 & 7.0$\pm$0.4 \\
\enddata
\tablenotetext{a}{Units: $\mathrm{erg\,s^{-1}\,cm^{-2}}$}
\end{deluxetable}
\vspace{-1em} 

As shown in Table~\ref{tab:1}, although the total fractional contributions from pure shock regions are comparable in both IFU datasets, the S7 WiFeS IFU data exhibit $\sim5-10\%$ smaller fractional contributions from the HII and AGN regions and a $\sim16-18\%$ larger contribution from fast shocks compared to the MUSE IFU data. {\color{black} This offset is caused by the differences in the seeing conditions and the spatial resolution between the two observations. The larger seeing in the S7 data ($\sim1.2\arcsec$ compared to $\sim0.66\arcsec$ in the MUSE observations) scatters light from the outer HII and AGN regions into surrounding spaxels that are excluded from the analysis due to low signal-to-noise ratio, resulting in both lower HII and AGN fractional contribution and lower total emission line fluxes in the S7 data. Additionally, the lower spatial resolution of the S7 data ($1\arcsec$ per pixel compared to $0.2\arcsec$ per pixel in the MUSE observations) may lead to the fast shock-dominated regions being spatially blended with adjacent AGN regions, where the bright fast-shock emission overshadows the surrounding AGN regions and lead to a higher fast shock fractional contribution.}

To summarize, both MUSE IFU data (with a physical resolution $\sim$40\,pc pixel$^{-1}$) and S7 IFU data (with a physical resolution $\sim$200\,pc pixel$^{-1}$) reveal similar structures in the center of NGC\,5728: a star-forming ring surrounding the galaxy center, an AGN-excited bicone structure origin from the galaxy center, and a fast shock-dominated region at the base of the AGN bicone structure. The consistent excitation mechanisms between MUSE IFU data and S7 IFU data suggest that IFU spectroscopy with coarser spatial resolution can still separate the mixing of HII, AGN, and shocks albeit with less detail if its spectral resolution is high enough (R$\gtrsim3000$) to perform multi-gaussian emission line fitting and accurate determination of the [OIII] emission-line velocity dispersion for each excitation component.

\section{Discussion}\label{sec:disc}

This paper presents a new theoretical 3D diagram that incorporates the latest photoionization models for HII regions, AGN regions, and shock radiation. The diagram is designed to facilitate the separation of star formation, AGN activity, and shocks in active galaxies. This 3D diagram consists of emission line ratios [N~II]$\,\lambda6584$/H$\alpha$, [O~III]$\,\lambda5007$/H$\beta$, and emission line velocity dispersion. Combined with IFU data, this 3D diagram can identify the HII, AGN, and shock-dominated regions, as well as estimate the relative contributions of different excitation mechanisms in areas where multiple sources are present.

We apply the new 3D diagram to galaxy NGC\,5728 using both MUSE IFU data and S7 IFU data. We found that the central $\sim5\,$kpc region of NGC\,5728 contains gas clouds that are dominantly ionized by HII regions, AGN radiation, and fast shock radiation, respectively. We also found regions where gas clouds are excited by both the HII regions and AGN radiation, and areas where AGN radiation and fast shocks jointly contribute to the ionization. {\color{black}Compared to the HII–AGN separation achieved using only 2D BPT diagrams \citep{durre_agn_2018,shin_positive_2019-1}, the application of this theoretical 3D diagram reveals previously unidentified fast shock and pure shock dominated regions in NGC\,5728.}
 
We further study the distribution of gas clouds excited by different mechanisms on the 2D maps of the galaxy NGC\,5728. We find that HII-dominated spaxels located along a ring structure surrounding the galaxy center with a projected radius of $\sim1\,$kpc in the sky plane, indicating the presence of a star-forming ring in the center of NGC\,5728. This star-forming ring and AGN ionization cones were reported in \citet{durre_agn_2018} using MUSE IFU data, later confirmed by the similar structure seen in the CO (2-1) emission from ALMA data \citep{shin_positive_2019,shimizu_multiphase_2019}. 

We calculate the star formation rate (SFR) of the star-forming ring ($2''<r<6''$) using the H$\alpha$ emission line in the field of view, and the L(H$\alpha$)-SFR relationship in \citet{robert_c_kennicutt_global_1998}. We corrected the observed H$\alpha$ emission following the steps in \citet{vogt_galaxy_2013} using $R_V^{A}=4.5$ and adopted D=40.3 Mpc from \citep{shimizu_multiphase_2019} as the distance of NGC\,5728. Using the corrected H$\alpha$ luminosity, we obtain SFR=1.17$\,\pm\,0.05\,M_{\odot}\,\rm yr^{-1}$ from MUSE data and SFR=0.57$\,\pm\,0.04\,M_{\odot}\,\rm yr^{-1}$ from the S7 data. We note, however, that the HII-region contribution in the S7 data is likely underestimated due to contamination from adjacent Seyfert-dominated and fast-shock dominated emissions. These SFRs are smaller than the SFR=1.38$\,\pm\,0.03$ $M_{\odot}\,\rm yr^{-1}$ estimated in \citet{davies_dissecting_2016} using S7 data because our measurements also exclude the contribution from shocks to the H$\alpha$ emission. Our estimated SFRs are also lower than the SFR of 1.48$\pm$0.01 $M_{\odot}\,\rm yr^{-1}$ derived from the 8$\mu m$ luminosity within the S7 field-of-view by \citet{davies_dissecting_2016}. This discrepancy is expected, given that dust can obscure all optical lines in some regions. 

Compared to the global SFR of 2.1 $M_{\odot}\,\rm yr^{-1}$ derived using {\it{Herschel}} photometry and SED fitting in \citet{shimizu_herschel_2017}, our result implies that the central star-forming ring contributes approximately $\sim50\%$ of the total star formation activity in NGC\,5728. Additionally, \citet{shimizu_multiphase_2019} reported evidence of gas inflow occurring at the intersection between the ring and dust lanes and estimated an inflow rate of 1 $M_{\odot}\,\rm yr^{-1}$. This inflow rate is comparable to the SFR we derived from the MUSE observations.


The AGN-dominated spaxels are distributed within a bicone structure extending $\sim2$\,kpc radially from the nucleus to the southeast and the northwest direction, indicating the presence of AGN ionization cones. {\color{black} Using the [O~III]/[O~I] AGN ionization parameter $\log(U)$ diagnostics and the R(N2,S2,H$\alpha$) AGN metallicity diagnostics in \citet{zhu_theoretical_2024}, we estimate the mean ionization parameter and gas metallicity of spaxels that have $f_{\rm AGN}>50\%$. For MUSE IFU data, the average parameters are $\log(U)=-2.9$ and 12+$\log(\rm O/H)=9.0$ assuming the `NHlow' scaling relation for AGN model, and $\log(U)=-3.0$ and 12+$\log(\rm O/H)=8.7$ assuming the `NHhigh' scaling relation for AGN model. The [O~I] line was not observed with sufficient S/N ratio to meet our selection criteria in S7 data. We instead use the [O~III]/[O~II] ratio to estimate the AGN ionization parameter for S7 observation. The average ionization parameter and metallicity in S7 data are $\log(U)=-3.0$ and 12+$\log(\rm O/H)=9.0$ for `NHlow' AGN model, and $\log(U)=-3.0$ and 12+$\log(\rm O/H)=8.7$ for `NHhigh' AGN model. 

We also constrain the black hole properties using the optical $E_{peak}$ diagnostic H$\beta$/[He~II]$\lambda$4686 ratio in \citet{zhu_theoretical_2024}. Only S7 data is available for this measurement because the [He~II]$\lambda$4686 emission line falls out of the observed wavelength range in MUSE data ($4750\lesssim\lambda\lesssim9360$). The average $E_{peak}$ of the AGN radiation field in NGC\,5728 is $\log(E_{peak}/\rm keV)=-1.86\pm0.17$ using `NHlow' AGN model and $\log(E_{peak}/\rm keV)=-1.84\pm0.18$ using `NHhigh' AGN model. Using the relationship between black hole mass ($M_{\rm BH}$), Eddington accretion rate ($f_{\rm Edd}$), and $E_{peak}$ in the Figure 3 of \citet{thomas_physically_2016}, our $E_{peak}$ measurements are consistent with the black hole mass $M_{\rm BH}=3.4\times10^7M_{\odot}$ and Eddington accretion rate $f_{\rm Edd}=3.3^{+7.3}_{-2.6}\times10^{-2}$ measured in \citet{durre_agn_2019}.}

The AGN ionization cones in NGC\,5728 were reported in \citet{durre_agn_2018,shin_positive_2019,shimizu_multiphase_2019} using optical MUSE IFU data, \citet{falcao_deep_2023} using deep \textit{Chandra} X-ray observations, and \citet{davies_gatos_2024} using JWST/MIRI MRS data. In addition to the AGN ionization, \citet{falcao_deep_2023} found that thermal emission from shocked gas also contributes to the X-ray spectra in the bicone region. The excitation mechanism 2D maps in our work also reveal a small fraction ($\sim20-40\%$) of shock contribution in Figure~\ref{fig:4-21}.

The fast shock-dominated spaxels are concentrated at the nuclear center with $\sim$ 1\, kpc length, and $\sim$80\,pc width, which is also at the base of the AGN-ionized bicone structure. The presence of fast shock at the base of AGN-ionized bicone suggests that gas in the regions is experiencing violent motion, {\color{black} with shock velocity $V_s\gtrsim300\,$km/s. These nuclear fast shocks in NGC\,5728 are likely associated with the radio jet presented in the AGN-ionized bicone direction. Simulations of the jet-ISM interaction in the galaxy with a central gas disc suggest that a relatively weak radio jet launch from the disc center can result in shock fronts along the disc direction \citep{mukherjee_relativistic_2018,meenakshi_modelling_2022}. 


At the fast shock-dominated region in NGC\,5728,} \citet{davies_gatos_2024} used JWST/MIRI MRS data and found an edge-on molecular disk with $\sim200$\,pc length. In addition, they also found that the power-law temperature distribution of the molecular gas can be well explained by the excitation from pure shocks with $V_s\approx30$\,km/s in dense ($10^5$\,cm$^{-3}$) gas. The coincidence of fast shocks in the ionized gas and pure shocks in the molecular gas in the galaxy nuclear region raises some interesting questions. Do the shocks have the same origins? If so, does this phenomenon suggest the stratification of different gas phases? Does the presence of nucleus shock relate to the accreting process of the nuclear accretion disk? 

Studies using the deep \textit{Chandra} X-ray observations found extended diffuse soft X-ray emission ($<$3\,keV) $\sim$1.4\,kpc in the cross-cone direction \citep{falcao_deep_2023}, indicating a mix of photoionization and shocked gas emission. This is expected in the fast shock regions, which include pure shock emission and photoionization from the precursor gas. Comparison with high-resolution observations at radio wavelengths is needed to gain further insight into the central accretion process of NGC\,5728. 

​In the separation of MUSE IFU data, we identify two distinct groups of spaxels dominated by pure shock emissions. The first group forms a shell encircling the fast shock-dominated nuclear region, while the second group is located further from the galaxy center ($r\approx1\,$kpc) and adjacent to regions dominated by HII emissions. The presence of pure shocks in proximity to fast shocks aligns with the anisotropic nature of shock propagation. Specifically, shock precursors emerge ahead of the shock front, whereas the regions trailing the shock front exhibit pure shock emissions. The second group of pure shock-dominated spaxels suggests mixing of star formation and shocks, indicative of stellar feedback mechanisms. Studies with higher spatial resolution observations are required to gain better insight into these regions.

We also compared the excitation mechanism between the MUSE and S7 IFU data. MUSE IFU, with a pixel size of 0.2$''$/pixel, a seeing of approximately 0.66$''$, and a field of view (FoV) of 1$' \times 1'$, clearly reveals a star-forming ring, an AGN ionized-bicone structure, a fast shock-dominated disk-shaped region at the base of the AGN ionized bicone, perpendicular to the bicone direction, {\color{black}and two group of pure shocks located near fast shock and star-forming regions, respectively.} Despite the coarser pixel size of 1$''$/pixel, a seeing of around 1.2$''$, and a smaller FoV of 38$'' \times 25 ''$, the excitation mechanism from the S7 IFU data still captures the presence of a star-forming arc, a bicone-shaped AGN outflow, a fast shock-dominated region at the base of the bicone, and some pure shock-dominated regions at $r\approx1-2\,$kpc near the star-forming regions.

The fact that coarser IFU data can reveal similar structures to those observed in higher spatially resolved IFU data is encouraging. This suggests that the methods used in this study could be applied to a large sample of galaxies where IFU data with $\sim$1$''$/pixel spatial resolution is available, such as in the SAMI and MaNGA surveys. {\color{black}In addition, the theoretical 3D diagnostic diagram developed in this work can be extended to high-redshift observations performed by high spatial-resolution IFU (i.e., JWST). For example, the JWST/NIRSpec IFU offers a spatial resolution of approximately 0.1$\arcsec$, corresponding to a physical scale of $\sim$0.4 kpc at $z = 1$, which is comparable to the physical spatial resolution achieved by ground-based IFU observations in the local universe.}

\section{Acknowledgement}

We thank the anonymous referee for thoughtful and useful comments, which have significantly improved this paper. PZ would like to thank Bethan James, Tim Heckman, Matilde Mingozzi, Lars Hernquist, and Pepi Fabbiano for their helpful suggestions in the early phase of this work. PZ also thanks Anna Trindade Falcao for useful discussion about the X-ray properties of NGC\,5728. PZ also thanks Pepi Fabbiano for helpful suggestions to improve this manuscript. Parts of this research were conducted by the Australian Research Council Centre of Excellence for All Sky Astrophysics in 3 Dimensions (ASTRO 3D), through project number CE170100013. KG is supported by the Australian Research Council through the Discovery Early Career Researcher Award (DECRA) Fellowship (project number DE220100766) funded by the Australian Government.

\bibliography{Paper3.bib}{}
\bibliographystyle{aasjournal}

\end{document}